# The Pulsar Wind Nebula Around PSR B1853+01 in the Supernova Remnant W 44

D. A. Frail

National Radio Astronomy Observatory, P.O. Box 0, Socorro, NM 87801

E. B. Giacani

Instituto de Astronomía y Física del Espacio, Buenos Aires, Argentina

W. M. Goss

National Radio Astronomy Observatory, P.O. Box 0, Socorro, NM 87801

and

G. Dubner

Instituto de Astronomía y Física del Espacio, Buenos Aires, Argentina

## ABSTRACT

We present radio observations of a region in the vicinity of the young pulsar PSR B1853+01 in the supernova remnant W 44. The pulsar is located at the apex of an extended feature with cometary morphology. We argue on the basis of its morphology and its spectral index and polarization properties that this is a synchrotron nebula produced by the spin down energy of the pulsar. The geometry and physical parameters of this pulsar-powered nebula and W 44 are used to derive three different measures of the pulsar's transverse velocity. A range of estimates between 315 and 470 km s$^{-1}$ are derived, resulting in a typical value of 375 km s$^{-1}$. The observed synchrotron spectrum from radio to X-ray wavelengths is used to put constraints on the energetics of the nebula and to derive the parameters of the pulsar wind.

*Subject headings:* Pulsars: individual (PSR B1853+01) – supernova remnants — ISM: individual (W 44)

## 1. Introduction

W 44 (G 34.7−0.4) belongs to a small group of supernova remnants in our Galaxy which contain a young, active pulsar. At radio wavelengths W 44 has a shell-type morphology, about 30′ in diameter, with the bulk of its radio emission concentrated in a series of knots and filaments (Jones, Smith & Angellini 1993). Its X-ray emission, mostly thermal in origin, is confined within the radio shell and is centrally-peaked (Smith et al. 1985, Rho et al. 1994). Although PSR B1853+01 is offset nearly 9′ south of the center of W 44, the similarities in their age and distance make a compelling case for a real, physical association between them (Wolszczan, Cordes & Dewey 1991). PSR B1853+01 has a period P of 267 ms and a period derivative $\dot{P}$ of $208\times10^{-15}$ s s$^{-1}$, implying a surface dipole field B$_o$ of $7.6\times10^{12}$ G and rotational energy loss rate $\dot{E}$ of $4.3\times10^{35}$ erg s$^{-1}$ (Wolszczan et al. 1991). Its small characteristic spindown age $\tau_c$ of only 20,000 yrs makes PSR B1853+01 one of the 10 youngest pulsars in our Galaxy (Frail, Goss & Whiteoak 1994).

It is widely thought that young pulsars efficiently transfer their rotational spindown energy $\dot{E}$ into a relativistic wind with some mean Lorentz factor $\gamma$ (Michel 1969, Rees & Gunn 1974, Kennel et al. 1983). The energy in the wind is divided between magnetic and electric flux (i.e. Poynting flux) and particle energy flux, parametrized by the ratio $\sigma$. From current models of pulsar magnetospheres it is expected that the wind will be magnetically dominated ($\sigma \gg 1$) out to at least the light cylinder radius (Arons 1992), but self-consistent models of the Crab nebula and pulsar require that $\sigma \ll 1$ further out (Kennel & Coroniti 1984, Emmering & Chevalier 1987). Several solutions have been proposed to explain this discrepancy (Begelman & Li 1994, Michel 1994, Melatos & Melrose 1996), including that by Coroniti (1990), who invoked magnetic reconnection from field reversals to lower $\sigma$ by taking energy from the B-field to accelerate particles. It is not known how common this behavior might be, since apart from the Crab pulsar and PSR B1957+20 (Kulkarni et al. 1992), there are very few measurements of $\sigma$. Another wind parameter of interest is the particle composition, expressed as k where k$\equiv$U$_i$/U$_e$, a ratio between the energy density in electrons U$_e$ versus all other particles U$_i$ (including positrons, protons and heavy ions). While most models assume an electron-positron pair plasma (k=1), an electron-ion plasma (k$\gg$1) is not ruled out, and may in fact be required to produce the observed power law distribution of the post-shock particles (Hoshino et al. 1992, Gallant et al. 1992).

The best way to constrain the wind parameters is through the study of synchrotron nebulae which appear as diffuse sources in the vicinity of some pulsars. The Crab nebula and other "plerions" are the best-known examples (Weiler & Sramek 1988). However, there are radio "wisps" and similar elongated structures, including the radio emission near the Vela pulsar (Bietenholz, Frail & Hankins 1991), the Crab pulsar (Bietenholz & Kronberg 1992), the CTB 80 pulsar (Strom 1987), the X-ray point source in 3C58 (Frail & Moffett 1993), and the source surrounding PSR B1757−24 (Frail & Kulkarni 1991). These features are thought to originate at the termination shock from a pulsar wind, formed where the wind first interacts with the surrounding medium. Thus the investigation of pulsar wind nebulae, their morphology, polarization structure, and spectral and temporal behavior, can put constraints on existing theoretical models.

Recently, Jones et al. (1993) reported a detailed X-ray and radio study of W44. In their 20-cm VLA radio images, they claim the detection of two discrete sources in the vicinity of the pulsar: weak extended radio emission at 20 cm, 12″ north of the pulsar with a flux of 2 mJy, and a "diamond-shaped" nebula 1.5 north, with a flux of 60 mJy. Jones et al. (1993) proposed that this diamond-shaped radio feature was a pulsar-powered synchrotron nebula. However, doubts remained about this identification because of the large offset between the pulsar and the nebula and the uncertain nature of the extended emission 12″ away from the pulsar. The present study was undertaken to address these difficulties by providing an improved position for PSR B1853+01 and imaging the region around the pulsar at a variety of radio frequencies.

## 2. Observations and Results

The interferometric position of the pulsar was determined from two datasets; the first was archival data taken on 1990 June 18 with the VLA in its BnA array configuration, and the second was on 1994 June 12 in the B array. Both observations used the right circular hands of polarization at 1465 MHz and 1515 MHz. In B array data there is only one radio source within a radius of 45″ of the nominal timing position from Wolszczan et al. (1991). A fit to the peak gives $\alpha(1950)=18^h 53^m 38.33^s$, $\delta(1950)=+01° 09' 26.1''$. This position is approximately 4″ to the northwest of the timing position. Since the timing position is based on only 18 months of timing data, and the pulsar exhibits timing noise comparable to that of the Crab pulsar (Wolszczan et al. 1991) such a discrepancy is not unexpected. At the higher resolution of the BnA array data (beamwidth of 1″.6 versus 4″) the emission in this region appears resolved, with three equally bright compact features clustered at the location of the B array peak. It appears that during the BnA array observations PSR B1853+01 was anomalously weak ($\leq$0.3 mJy) and thus we cannot uniquely identify which of these peaks is the pulsar and which are partially resolved extended emission. While the formal errors on the B array fits are of order $\pm 0.15''$, we assign an error of $\pm 1.5''$ to reflect the uncertainty in correctly identifying the pulsar in the presence of extended emission.

The extended emission in the vicinity of the pulsar was imaged at 4860 MHz and 8440 MHz on 1995 May 20 with the VLA in its D array configuration. Two 50 MHz channels were measured at 6 cm and 3.6 cm in both parallel and cross hands of circular polarization. All data reduction and calibration was done following standard practice in use at the VLA. In Figure 1b we show the 8.4 GHz ($\lambda$=3.6 cm) image of the pulsar, and for perspective in Figure 1a we include a 1.4 GHz image showing the location of the pulsar in relation to the larger W44 remnant. The 1.4 GHz image is part of a comprehensive radio, X-ray and optical study of the W44 remnant (Giacani et al. 1996). Unlike a similar image at 1.4 GHz by Jones et al. (1993), care was taken to add the missing short order spacings using single dish data. Consequently the true nature of the emission around PSR B1853+01 is revealed.

With the addition of short spacings the two discrete sources seen by Jones et al. (1993), the "diamond" nebula and the extended emission 12″ north of the pulsar, form a single structure.

The morphology is even better defined in Fig. 1b. The pulsar is located at the apex of a cometary-shaped radio source 2′.5 in extent. A narrow "contrail" leads away from the pulsar to the northwest where it widens out and the continuum emission peaks. There is a "kink" in the nebula about 30″ from PSR B1853+01, that could be produced by density or temperature variations within W 44, and should be detectable as fluctuations in the X-ray emissivity. Similar axisymmetric structures have been seen at radio wavelengths including G 5.4−1.2 (Frail & Kulkarni 1991), G 359.2−0.8 (Yusef-Zadeh & Bally 1987, Predehl & Kulkarni 1994) and G 354.1+0.1 (Frail et al. 1994). Like the nebula in Fig. 1, these are best understood as synchrotron nebulae that arise as a result of the spin down of young, energetic pulsars.

Several other lines of evidence support the contention, first put forth by Jones et al. (1993), that this is a synchrotron nebula generated by the pulsar and not just a superposition of a filament from W 44 projected onto PSR B1853+01. The total flux density of the nebular emission around PSR B1853+01 is 287±126 mJy at 20 cm and 220±26 mJy at 3.6 cm. A radio spectrum between 330 MHz and 8.4 GHz using data from Giacani et al. (1996) is given in Figure 2. The larger error bars for the lower frequency radio data reflect the difficulty in separating this emission from its surroundings. At these frequencies this nebular emission is $< 10^{-3}$ of the total flux density of W 44 (Kovalenko, Pynzar' & Udal'tsov 1994a,b). We include an X-ray upper limit from the Ginga satellite determined by Harrus, Hughes & Helfand (1996) who also report on the detection of an extended, hard X-ray source near this flux level with the ASCA satellite. A weighted least-squares fit to the radio portion of the spectrum yields $\alpha_R = -0.12 \pm 0.04$ (where $S_\nu \propto \nu^{\alpha_R}$), substantially flatter than $\alpha_R = -0.33$ for the rest of W 44 (Kovalenko et al. 1994a,b). The emission is polarized with a degree of polarization at 3.6 cm and 6 cm of approximately 17±4%. Although W 44 exhibits a high degree of polarization of up to 20% at frequencies above 10 GHz (Kundu & Velusamy 1972), the emission is strongly depolarized at lower frequencies, dropping to a peak of 3% at 2.7 GHz (Velusamy & Kundu 1974). At 5 GHz (6 cm) we estimate from the data of Whiteoak & Gardner (1971) that the degree of polarization for W 44 in the vicinity of the pulsar is only 3.5%. At this same frequency our data shows that the polarized flux from the compact nebula stands out from the body of the surrounding emission, confirming its much higher degree of polarization compared to W 44 as a whole.

Thus the case for the synchrotron source around PSR B1853+01 being a pulsar-powered wind nebula (PWN) rests on its unusual cometary morphology with the pulsar at its apex, and the fact that its radio spectrum and degree of polarization distinguish it from the rest of W 44. In these latter two properties this PWN has much in common with other pulsar-powered nebula or "plerions" of which the Crab nebula is the prototype (Weiler & Sramek 1988). In the next section we use the observed properties of the PWN to put constraints on the pulsar parameters.

## 3. Discussion

The morphology of the PWN can be used to put independent constraints on both the magnitude and direction of the transverse velocity $V_{PSR}$ of PSR B1853+01. Wolszczan et al. (1991) and Frail et al. (1994) earlier derived $V_{PSR}$ of PSR B1853+01 on the basis of the pulsar's displacement from the geometric center of the remnant and its age. A better estimate of these can be made, since the birthplace of PSR B1853+01 can be located by three different methods. The geometric center of the radio continuum shell (Jones et al. 1993), the dynamical center of the HI shell (Koo & Heiles 1995), and the peak of the X-ray emission (ROSAT: Rho et al. 1994, Einstein: Seward 1990) all pinpoint the probable center and agree quite well, giving an angular displacement of PSR B1853+01 from this point of $8\rlap{.}''6\pm0\rlap{.}''5$. This point lies along a line that bisects the axis of symmetry of the PWN, suggesting that the pulsar originated from this location. It follows that the implied velocity of PSR B1853+01 is $V_{PSR} = 370\,d_3\,t_{20}^{-1}$ km s$^{-1}$ ($\pm 25$ km s$^{-1}$), where the age for W 44 of 20,000 $t_{20}$ years is from $\tau_c$ of PSR B1853+01, which is likely an upper limit.

Another estimate of the velocity of PSR B1853+01 is possible. For any supernova remnant we can express the shock velocity at a time t as $V_s = c_\circ\,R_s/t$, where $R_s$ is the radius and $c_\circ$ is a constant equal to 2/5 for a remnant in the Sedov phase (Shull et al. 1989). Similarly, the pulsar's velocity is $V_{PSR} = \beta\,R_s/t$, where $\beta$ is the fractional displacement of the pulsar from the center of the remnant. Combining these two equations we see that the pulsar velocity and shock velocity are related by $V_{PSR}=V_s\,\beta/c_\circ$. For PSR B1853+01 we estimate $\beta \simeq 0.5$, while Koo & Heiles (1995) derive $V_s$=330 km s$^{-1}$ for the radio continuum shell of W 44. If W 44 is near the Sedov phase (as favored by Koo & Heiles) then for $c_\circ \simeq 0.35$-$0.45$ we derive $V_{PSR}$=370-470 km s$^{-1}$.

A third way to estimate the pulsar velocity is through the pressure balance condition that must exist between the PWN and the hot gas interior to W 44. According to Rees & Gunn (1974) the relativistic pulsar wind flows outward until it terminates at a point $r_s$, where it comes into pressure equilibrium with the surrounding medium. For W 44 this pressure is given by

$$\frac{\dot{E}}{4\pi\,r_s^2 c} = 5.7 \times 10^{-10} \theta_s^2\,d_3^{-2}\ \mathrm{erg\ cm^{-3}}, \qquad (1)$$

where we have assumed an isotropic wind and $\theta_s$ is the angular radius (in arcsec) where the reverse shock is formed. This region appears unresolved in Fig. 1 but from our high resolution data we estimate that $\theta_s \leq 1''$. The distance of $3\,d_3$ kpc is well-established by HI absorption and 1720 MHz maser lines (Caswell et al. 1975, Claussen et al. 1996). Some fraction of $\dot{E}$ could be converted into high-energy radiation rather than particles and field energy in the wind but upper limits for the detection of pulsed emission from PSR B1853+01 with ROSAT (0.1-2.4 keV, Rho et al. 1994) and EGRET ($\geq$100 Mev, Thompson et al. 1994a) suggest that the production efficiency $\eta$ is low ($\leq 10\%$). The unusual cometary morphology leaves little doubt that the source of external pressure confining the flow is the ram pressure $\rho V_{PSR}^2$ due to the pulsar's motion through W 44,

$$\rho\,V_{PSR}^2 = 2.3 \times 10^{-10}\,n_\circ V_{100}^2\ \mathrm{erg\ cm^{-3}}. \qquad (2)$$

We have allowed for a 10% composition of He by mass in calculating the density $n_o$. $V_{100}$ is the pulsar velocity in units of 100 km s$^{-1}$. Equating eqns (1) and (2) and using $n_o$=0.25 cm$^{-3}$ near PSR B1853+01 from Jones et al. (1993) we derive $V_{PSR} \geq 315\,\theta_s^{-1} d_3^{-1}$ km s$^{-1}$.

Beyond $r_s$ the wind becomes sub-relativistic and is swept back and kept separate from the shocked external gas by a bow shock that leads ahead of the pulsar (Cheng 1983, Wang, Li & Begelman 1993). A synchrotron trail forms in a direction opposite to the pulsar's motion, confined by the external medium with a length determined by adiabatic and synchrotron losses. For the 2$\rlap{.}'$5 long PWN trailing behind PSR B1853+01 the thermal pressure from the hot X-ray emitting gas in W44 is approximately 6-8×10$^{-10}$ erg cm$^{-3}$ (Jones et al. 1993), comparable to the ram pressure in eqn. 2 and several orders of magnitude above the pressure in the general interstellar medium of $4\times 10^{-13}(n_o\,T_o/3000)$ erg cm$^{-3}$ (Kulkarni & Heiles 1988). This trail points in a northwest direction, supporting our contention made earlier, that the pulsar originated close to the geometric center of W44.

In summary, we find that all three independent methods yield an estimate for $V_{PSR}$ which agree remarkably well, suggesting that in the absence of proper motion measurements, inferred velocities are basically sound. We adopt a value $V_{PSR}$=375 km s$^{-1}$ for the transverse velocity of PSR B1853+01 and predict proper motion of order 25 mas yr$^{-1}$.

The observed synchrotron spectrum from radio to X-ray wavelengths (Fig. 2) can be used to put constraints on the energetics of the nebula (Helfand & Becker 1987, Salter et al. 1989) and the parameters of the pulsar wind discussed in §1 (i.e. $\gamma$, $\alpha$ and k). From the spectrum displayed in Fig. 2, it can be seen that the Ginga X-ray upper limit does not lie on a smooth extrapolation of the radio spectrum, implying that the spectrum must break at some $\nu_B$. Spectral breaks hold the key to understanding the physical conditions of the nebula and the pulsar wind. If the break results from synchrotron losses then, given the age of the source $t_{res}$, we can estimate the nebular field $B_n$. From $V_{PSR}$ and the linear size of the PWN, we derive $t_{res} \simeq 5700$ yrs and from the standard formulae in Pacholczyk (1970) $B_n$=1040 $\mu$G $(\nu_B/10^{12}$ Hz$)^{-1/3}(t_{res}/1000$ yrs$)^{-2/3}$. A Crab-like spectral index for the X-ray emission $\alpha_X = -1$ gives $\nu_B = 5.7 \times 10^{11}$ Hz and $B_n \simeq 400\,\mu$G. If $\alpha_X$ is steep, say $-1.5$, then $\nu_B = 1.2 \times 10^{14}$ Hz and $B_n \simeq 70\,\mu$G. An upper limit of $\alpha_X = -0.81$ is derived by setting $\nu_B$ to $10^{10}$ Hz, corresponding to the far end of the observed radio spectrum. In this case $B_n \simeq 1500\,\mu$G.

A lower limit to the value of $\nu_B$ can be derived by using other measures of the magnetic field. The maximum field reached in the pulsar wind is when it remains Poynting flux-dominated ($\sigma > 1$) out to the reverse shock. Then from eqn. 1 we can estimate the equivalent magnetic field $B_E = \sqrt{2\dot{E}/r_s^2 c}$ which would contain the energy density of the wind at $r_s$ (Thompson et al. 1994b). Similarly, if we assume that the magnetic field, which is $B_o$ at the surface of the neutron star at a radius R, falls off as a dipole out to the light cylinder radius $r_{lc} \equiv \Omega/c$ and is thereafter toroidal to $r_s$, then $B_s$=$B_o(R/r_{lc})^3(r_{lc}/r_s)$. Both values give $B_s \simeq 110\,\mu$G but it could be much lower if $\sigma$ decreases from $r_{lc}$ to $r_s$, as seems necessary for the Crab nebula (Kennel & Coroniti

1984, Emmering & Chevalier 1987). Some enhancement of the postshock $B_n$ is expected over $B_s$ but at least in the case of a strong relativistic shock, $B_n \leq 3B_s$ (Kennel & Coroniti 1984). From the values of $B_n$ derived earlier this allows us to limit $\nu_B \geq 10^{12}$ Hz.

It follows that the Lorentz factor $\gamma$ for the electrons responsible for the emission near $\nu_B$ is $\gamma \sim 10^5 (\nu_B/10^{12} \text{Hz})^{1/2} (B/100\,\mu\text{G})^{-1/2}$ (Pacholczyk 1970). Larger values of $\gamma$ and its evolution away from the pulsar are measurable in principle from high resolution X-ray observations. Integrating the spectrum from $10^7$ Hz to $\nu_B = 10^{12}$ Hz we derive a radio luminosity $L_R$ of $1.5 \times 10^{33} d_3$ ergs s$^{-1}$. Likewise from the Ginga data the X-ray luminosity $L_X$ is $1.3 \times 10^{33} d_3$ ergs s$^{-1}$. Both $L_R$ and $L_X$ are only $\sim 0.003\,\dot{E}$. The empirical relation of Seward & Wang (1988) between $L_X$ and $\dot{E}$ for PWN predicts $L_X \simeq 0.002\,\dot{E}$ for the nebula around PSR B1853+01. Adopting the equipartition condition (Pacholczyk 1970) we derive $B_{eq} \geq 60\,\mu\text{G}\,(1+k)^{2/7}$ ($\nu_B \geq 10^{12}$ Hz) and a minimum energy $E_{min} \geq 1.3 \times 10^{46}$ ergs $(1+k)^{4/7}$ for the PWN. It is not hard to show that for a dipole rotator the energy deposited by the pulsar into the nebula is $\dot{E} \times t_{res}\,[t_{20}/(t_{20}-t_{res})]$, and thus an upper limit on k can be derived by requiring $E_{min} \simeq 1.4\,\dot{E} \times t_{res} \simeq 10^{47}$ ergs. At $\nu_B = 10^{12}$ Hz k=30 but it falls to k=5 at $\nu_B = 10^{14}$ Hz, suggesting that, while ions may exist in the wind, they are evidently energetically less important than some models assume (Gallant & Arons 1994).

Finally we comment that if the magnetic field $B_n$ were better known (from $\nu_B$) then the energy density ratio between magnetic fields $U_B$ and particles $U_p$ in the nebula $\sigma_n \equiv U_B/U_p$ could be determined (since $\sigma_n \propto V\,B_n^{7/2}/(1+k)L_R$, where V is the nebular volume), leading ultimately to $\sigma$ of the wind. The constraints on the pulsar wind diagnostics deduced here will improve once a better measurement of $\nu_B$ is made and the data are interpreted with more complex models (e.g. Reynolds & Chevalier 1984, Reynolds & Chanan 1984) which will need to incorporate continuous particle/field injection into the cometary-shaped nebula produced by the pulsar's space motion.

The Very Large Array (VLA) is a facility of the National Science Foundation operated under cooperative agreement by Associated Universities, Inc. This research has made use of the Simbad database, operated at CDS, Strasbourg, France. DAF thanks S. Kulkarni, J. Navarro and G. Vasisht for useful discussions.

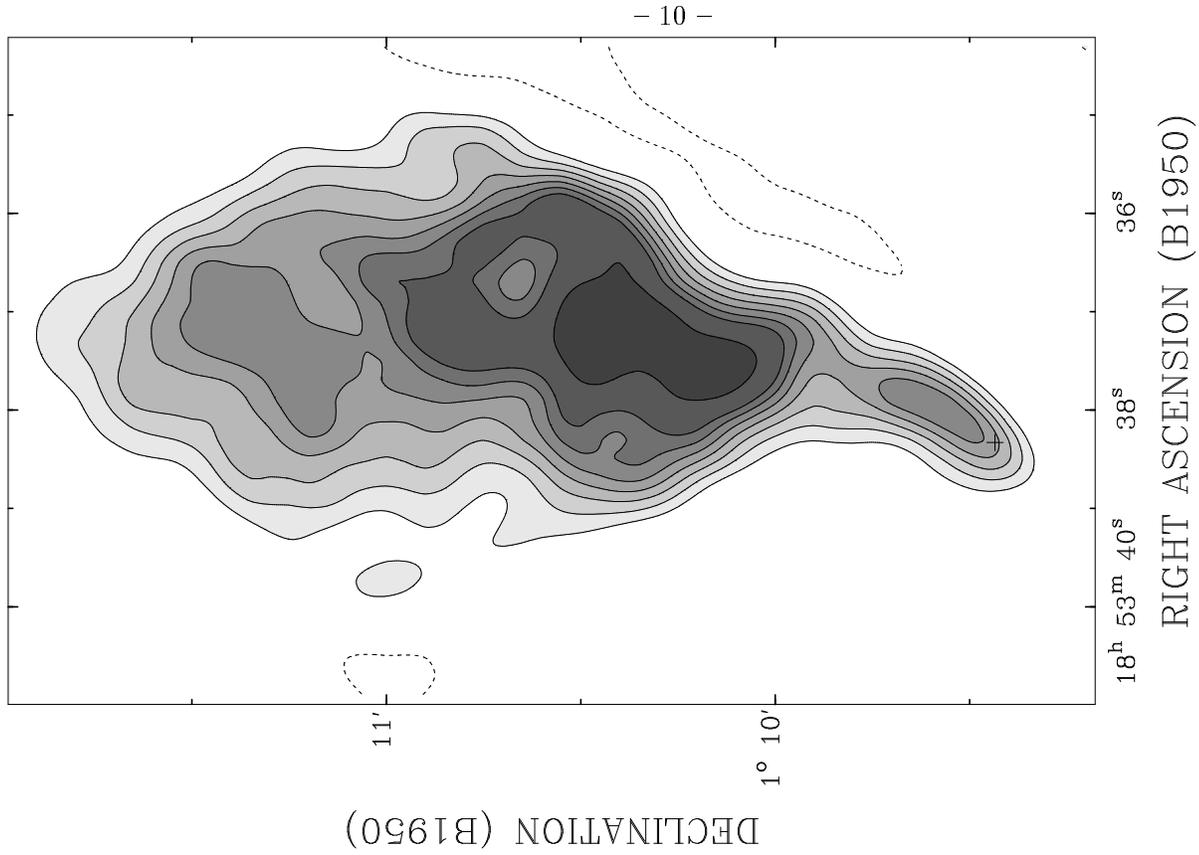
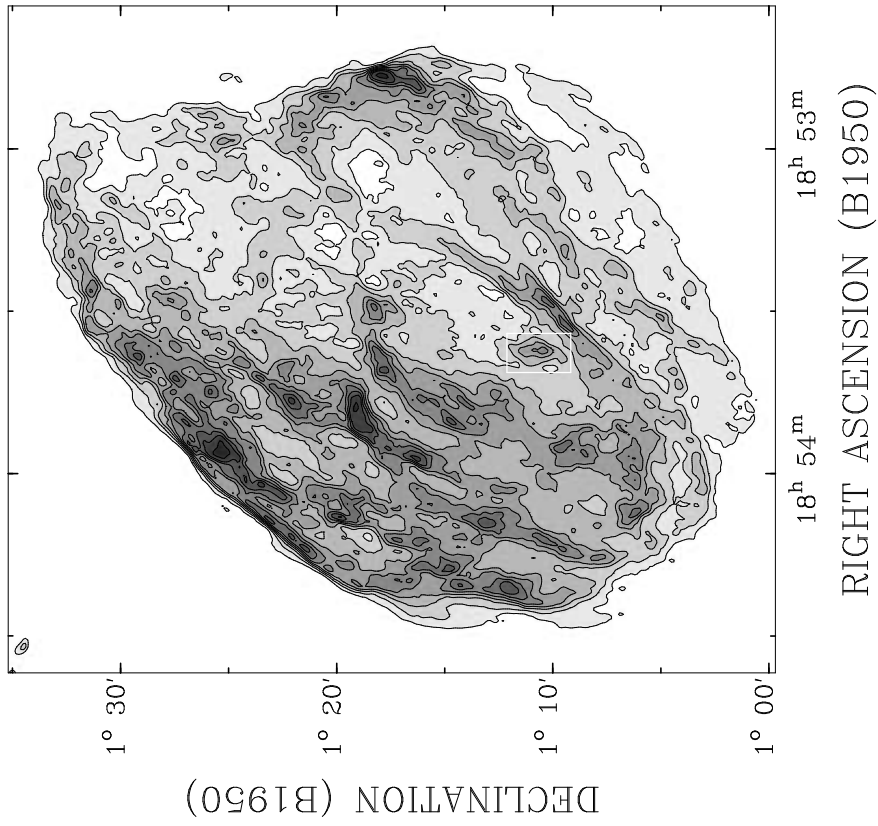

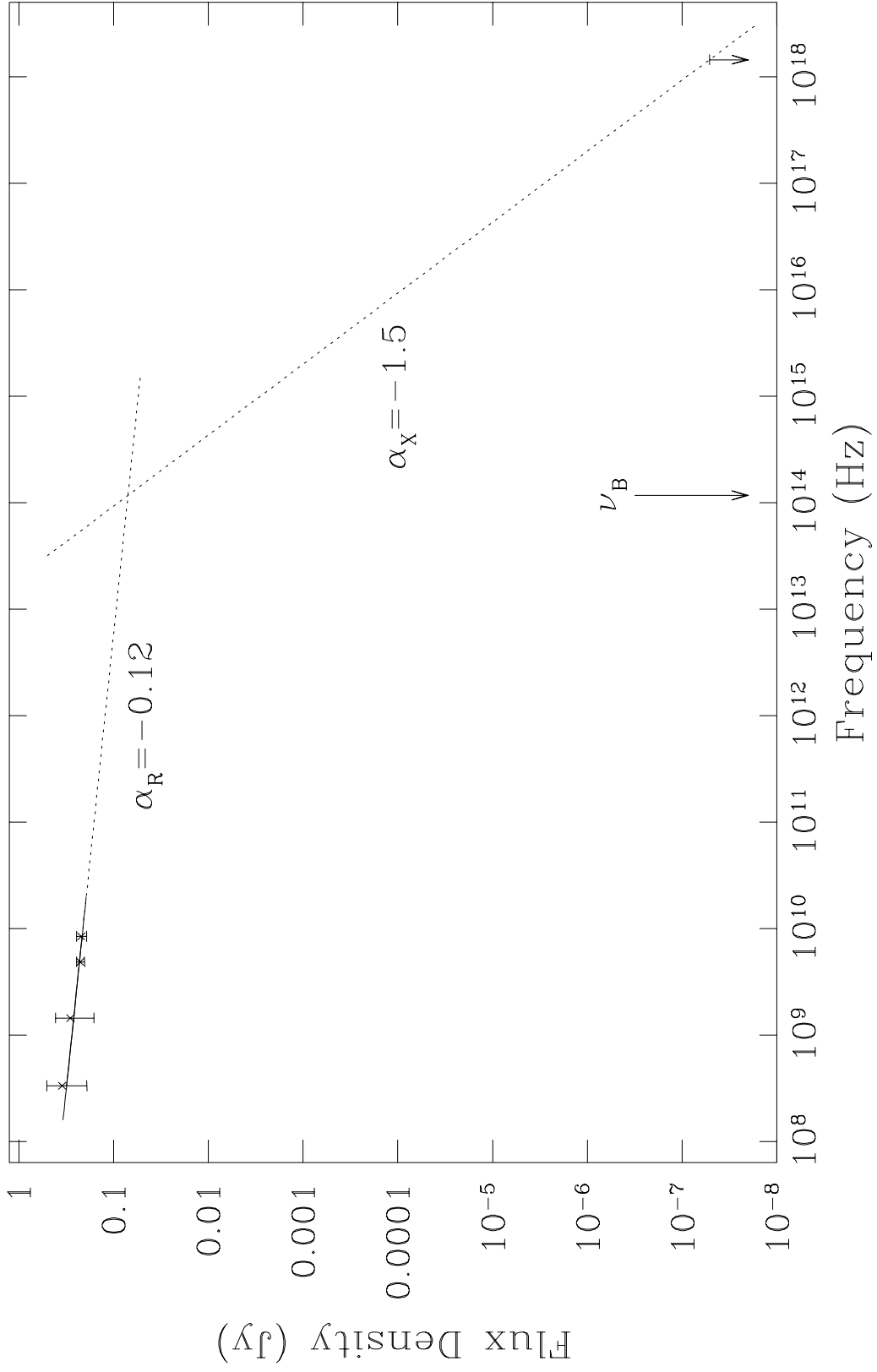

Fig. 1.— (a) A 20-cm radio continuum image of the supernova remnant W 44. The inset region is shown in more detail in (b) with a 3.6 cm image showing the pulsar wind nebula around PSR B1853+01. The position of the pulsar is indicated by a cross (+). Contour intervals are $-0.6$, 0.6, 1, 1.4, 1.8, 2.2, 2.6, 3, 4 mJy beam$^{-1}$. The rms noise is 0.2 mJy beam$^{-1}$. The synthesized beam size is $8''\!.9 \times 7''\!.8$, P.A.$=-24°$.

Fig. 2.— The spectrum of the pulsar wind nebula around PSR B1853+01 over the range from $10^8$ Hz to $10^{18}$ Hz. Solid line with slope $\alpha_R$ in the radio regime is a weighted least-squares fit to the data points discussed in the text. The X-ray point is an upper limit from Ginga data (see text). A Crab-like power-law spectral index $\alpha_X = -1.5$ is drawn through the X-ray data point and intersects the extrapolation of the radio spectrum at a break frequency $\nu_B$.